\documentclass[pra,superscriptaddress,showpacs,twocolumn,10pt]{revtex4}
\usepackage{amssymb}
\usepackage{graphicx}

\def\>{\rangle}
\def\<{\langle}

\newcommand{\tr}{{\rm Tr}}
\newcommand{\supp}{{\rm supp}}
\newcommand{\diag}{{\rm diag}}

\begin{document}

\title{Condition and capability of quantum state separation}
\author{Yuan Feng}
\email{feng-y@tsinghua.edu.cn}
\author{Runyao Duan}
\email{dry02@mails.tsinghua.edu.cn}
\author{Zhengfeng Ji}
\email{jizhengfeng98@mails.tsinghua.edu.cn}

\affiliation{State Key Laboratory of Intelligent Technology and
Systems, Department of Computer Science and Technology, Tsinghua
University, Beijing, China, 100084}

\date{\today}

\begin{abstract}
The linearity of quantum operations puts many fundamental
constraints on the information processing tasks we can achieve on
a quantum system whose state is not exactly known, just as we
observe in quantum cloning and quantum discrimination. In this
paper we show that in a probabilistic manner, linearity is in fact
the only one that restricts the physically realizable tasks. To be
specific, if a system is prepared in a state secretly chosen from
a linearly independent pure state set, then any quantum state
separation can be physically realized with a positive probability.
Furthermore, we derive a lower bound on the average failure
probability of any quantum state separation.
\end{abstract}


\maketitle

\newtheorem{corollary}{Corollary}
\newtheorem{definition}{Definition}
\newtheorem{example}{Example}
\newtheorem{lemma}{Lemma}
\newtheorem{proposition}{Proposition}
\newtheorem{theorem}{Theorem}
\newtheorem{fact}{Fact}
\newtheorem{property}{Property}
\renewcommand{\choose}[2]{{{#1}\atopwithdelims(){#2}}}

\date{\today}
\section{Introduction}
A fundamental difference between quantum mechanics and the
classical correspondence is that in the former, a system can be
not only in a basis state but also in a state which is a linear
combination, or `superposition', of different basis states.
Quantum computation and quantum information processing benefit
extremely from superposition since performing a quantum operation
on a superposition is equivalent to performing the same operation
synchronously on all of the basis states constituting this
superposition. One of the most famous examples is Shor's quantum
factoring algorithm \cite{SH94}. On the other hand, however, the
existence of superposition in quantum mechanics also puts many
constraints on the physically realizable information processing
tasks, when we have only limited information about the original
state of the system that we are concerned with. Take quantum
cloning, perhaps the most fundamental task in quantum computation
and quantum information processing, as an example. When the state
to be cloned is thoroughly known, it can be perfectly cloned by
using a state-dependent cloning machine (In fact, since the state
is known, we can prepare as many copies of it as needed. The
reason behind it is in fact that classical information can be
cloned arbitrarily). Here and in the rest of this paper, by
`perfectly' we mean the information processing task is realized
with certainty and without any approximation or error. Suppose
further we want to build a universal cloning machine for different
pure states, then only if these states are linearly independent
that a desired exact cloning machine exists even in a
probabilistic manner \cite{DG98}. The possibility to obliviously
clone states from a linearly dependent set is forbidden by the
linearity of quantum operations. Another result in Ref.
\cite{DG98} which receives less attention than it deserves is the
converse of the above statement. That is, when the possible states
of the original system are linearly independent, then
$1\rightarrow N$ probabilistic cloning is possible for any $N\geq
1$. In this paper, we generalize this result to show that the
linear independency of the original states is enough to make any
information processing tasks possible in a probabilistic manner.

Another fundamental task in quantum computation and quantum
information processing is quantum discrimination. Given that the
system of interest is prepared in one of some possible states, the
purpose of discrimination is to tell which state the system is
actually in. Rather surprisingly, these two seemingly, at least at
first glance, very different tasks are closely related. A quantum
system can be perfectly cloned \cite{WZ82,DI82} (resp. perfectly
discriminated \cite{HE76}) if and only if the possible states of
the system are orthogonal; and it can be conclusively cloned
\cite{DG98} (resp. unambiguously discriminated \cite{DG98,CH98})
if and only if the possible states are linearly independent.
Furthermore, Duan and Guo \cite{DG98} pointed out that exact
$1\rightarrow \infty$ cloning and unambiguous discrimination can
be simulated by each other; a more delicate and quantitative
connection between these two tasks was investigated in Ref.
\cite{CB98}.

Motivated by this connection, Chefles and Barnett proposed a
generalized way, namely quantum separation, to deal with quantum
exact cloning and quantum unambiguous discrimination uniformly
\cite{CB98}. To be specific, suppose a quantum system is prepared
in one of the two states $|\psi_1\>$ and $|\psi_2\>$ but we do not
know exactly which one. A quantum separation performed on this
system then leads, generally in a probabilistic but conclusive
manner, the system into $|\psi'_i\>$ provided that originally it
is in the state $|\psi_i\>$, for $i=1,2$. In their paper, Chefles
and Barnett put a constraint that the desired states $|\psi'_1\>$
and $|\psi'_2\>$ should satisfy the condition that
\begin{equation}\label{constraint}
|\<\psi'_1|\psi'_2\>|\leq |\<\psi_1|\psi_2\>|,
\end{equation}
just as in the cases of exact cloning and unambiguous
discrimination. That is also why they called this process
`separation' since decrease of the inner product means that these
two states become more distinct or separable. In the present
paper, we generalize this concept in two ways. First, we get rid
of the constraint in Eq.(\ref{constraint}) to consider more
general physical processes, although we still use the term
`separation' for convenience. Second, we generalize separation to
the case of multiple mixed states.

To be specific, we give the formal definition of quantum
separation as follows. Suppose a quantum system is prepared in a
state secretly chosen from $\rho_1,\dots,\rho_n$. A quantum
separation is a physically realizable process which, generally in
a probabilistic but conclusive manner, leads $\rho_i$ to $\rho'_i$
for some quantum states $\rho'_1,\dots,\rho'_n$. Recall that any
physically realizable process is completely positive and trace
preserving, and so can be represented by Kraus operator-sum form
\cite{KR83}. That is, there exist quantum operators $A_{Sk},
A_{Fk}$ such that
\begin{equation}\label{1}
A_{Sk}\rho_i A_{Sk}^\dag=s_{ik}\rho'_i,
\end{equation}
\begin{equation}\label{2}
A_{Fk}\rho_iA_{Fk}^\dag=f_{ik}\sigma_{ik}
\end{equation}
for some nonnegative real numbers $s_{ik}$ and $f_{ik}$, and mixed
states $\sigma_{ik}$, where $i=1,\dots, n$. Here the subscript $S$
and $F$ denote success and failure, respectively. Intuitively,
Eq.(\ref{1}) means that if the separation succeeds, the system
evolves into $\rho'_i$ provided that it is originally in the state
$\rho_i$. Notice that there may be more than one operator, indexed
by $k$, corresponding to successfully separating $\rho_i$ or
getting an inconclusive result. By appending the shorter group
with zero operators, we can assume that the range of $k$ is taken
the same for success and failure. Furthermore, these operators
should satisfy the completeness relation
\begin{equation}\label{identity}
\sum_{k}(A_{Sk}^\dag A_{Sk} + A_{Fk}^\dag A_{Fk}) = I.
\end{equation}
Here $I$ is the identity operator.

Since no constraints are put on the output states in the general
framework, we can in fact represent any oblivious computation and
information process by quantum state separation. To see the power
of this framework more explicitly, let us examine some special
cases. It is easy to check that exact $1\rightarrow N$ cloning is
a special case of quantum separation by letting the desired state
$\rho'_i$ be $\rho_i^{\otimes N}$ while unambiguous discrimination
is the case when all $\rho'_i$ are orthogonal such that there
exists a quantum measurement which can further discriminate them
perfectly. Furthermore, suppose all $\rho_i$ lie in a Hilbert
space $\mathcal{H}$. The $1\rightarrow N$ mixed state broadcasting
\cite{BC96} can be involved in the general framework of quantum
state separation by requiring that each $\rho'_i$ lies in the
Hilbert space $\mathcal{H}^{\otimes N}$ and the reduced density
matrices of $\rho'_i$ obtained by tracing over any $N-1$
subsystems equal to $\rho_i$. Note also that unambiguous filtering
\cite{TB03}, unambiguous comparison \cite{BC03}, and unambiguous
subset discrimination \cite{ZY02} are all special cases of
unambiguous discrimination between mixed states, which has
received much attention in recent years [13-19]. By considering
quantum state separation, we can deal with all these information
processing processes in a uniform and more general way.

The aim of this paper is to examine the conditions and the
capability of quantum information processing in the framework of
state separation. In Sec. II, we show that in order to physically
realize a universal and conclusive information processing task on
an unknown system, linearity is in fact the only constraint. In
other words, when the possible states of the unknown system are
linearly independent, then any separation with any output states
is possible. In Sec. III, we derive a lower bound on the average
failure probability of any physically realizable quantum
separation, when the mixed state case is considered.

\section{Conditions of state separation}

In this section, we derive some necessary and sufficient
conditions for quantum separation to be physically realizable.
First, when the final states are specified, we have the following
theorem for the pure state case.

\begin{theorem}
Given two sets of pure states $|\psi_1\>,\dots,|\psi_n\>$ and
$|\psi'_1\>,\dots,|\psi'_n\>$. There exists a quantum separation
which can lead $|\psi_i\>$ to $|\psi'_i\>$ if and only if
\begin{equation}\label{large}
X-\sqrt{\Gamma} X' \sqrt{\Gamma}\geq 0
\end{equation}
for some positive definite diagonal matrix
$\Gamma=\diag(\gamma_1,\dots,\gamma_n)$, where $n\times n$
matrices $X=[\<\psi_i|\psi_j\>]$ and $X'=[\<\psi'_i|\psi'_j\>]$.
Here by $M \geq 0$ we mean that the matrix $M$ is positive
semidefinite, i.e., for any $n$-dimensional complex vector
$\alpha$, $\alpha M \alpha^\dagger \geq 0$.
\end{theorem}

To prove this theorem, we introduce first a lemma proven in Ref.
\cite{DG98}:
\begin{lemma}
For any two sets of pure states $|\psi_1\>,\dots,|\psi_n\>$ and
$|\psi'_1\>,\dots,|\psi'_n\>$, if
\begin{equation}
\<\psi_i|\psi_j\>=\<\psi'_i|\psi'_j\>
\end{equation}
for any $i,j=1,\dots,n$, then there exits a unitary operator $U$
such that $U|\psi_i\>=|\psi'_i\>$.
\end{lemma}
We learn from this lemma that in pure state case, the only thing
determining whether or not there exists a unitary evolution
between two sets of states is the inner products of all pairs of
states from the same set. This is a remarkable property of pure
state evolution. When mixed states are considered, things become
more complicated and many more facts other than fidelities between
different states must be involved to determine the existence of
such a unitary transformation. That is also why we consider only
pure state case here.

Having the above lemma as a tool, we can prove Theorem 1 as
follows:

{\it Proof of Theorem 1.} By definition, there exist quantum
operators $A_{Sk}$ and $A_{Fk}$ satisfying Eq.(\ref{identity})
such that
\begin{equation}\label{pure1}
A_{Sk}|\psi_i\>= \sqrt{s_{ik}}|\psi'_i\>
\end{equation}
\begin{equation}\label{pure2}
A_{Fk}|\psi_i\>=\sqrt{f_{ik}}|\phi_{ik}\>,
\end{equation}
for some state $|\phi_{ik}\>$, where $0<s_{ik}\leq 1$ and $0\leq
f_{ik}< 1$. For any $n$-dimensional complex vector
$\alpha=(\alpha_1,\dots,\alpha_n)$, let $|\Psi\>=\sum_{i=1}^n
\alpha_i |\psi_i\>$. Notice that $A_{Fk}^\dagger A_{Fk}$ is
positive semidefinite for any $k$. It follows that
\begin{equation}\label{positive}
\begin{array}{rl}
0 & \displaystyle\leq \<\Psi|\sum_k A_{Fk}^\dagger A_{Fk}|\Psi\>\\
\\
&=\displaystyle\<\Psi|I-\sum_k A_{Sk}^\dagger A_{Sk}|\Psi\>\\
\\
&=\displaystyle\<\Psi|\Psi\>-\sum_k \<\Psi|A_{Sk}^\dagger A_{Sk}|\Psi\>\\
\\
&=\displaystyle\sum_{i,j}\alpha_i^*\alpha_j\<\psi_i|\psi_j\>-\sum_k \sum_{i,j}\alpha_i^*\alpha_j\sqrt{s_{ik}s_{jk}}\<\psi'_i|\psi'_j\>\\
\\
&=\displaystyle\alpha X  \alpha^\dagger -  \alpha \sum_k
\sqrt{S_k}X'\sqrt{S_k}\alpha^\dagger\\
\\
&\displaystyle\leq \alpha X  \alpha^\dagger -  \alpha
\sqrt{S_1}X'\sqrt{S_1}\alpha^\dagger.
\end{array}
\end{equation}
Here, $S_k=\diag(s_{1k},\dots,s_{nk})$ are $n\times n$ diagonal
matrices. The last line of Eq.(\ref{positive}) follows from the
fact that for any $k$, $\sqrt{S_k}X'\sqrt{S_k}$ is positive
semidefinite. From the arbitrariness of $\alpha$, we derive that
\begin{equation}
X  -  \sqrt{S_1}X'\sqrt{S_1} \geq 0,
\end{equation}
which completes the proof of the necessity part.

The proof of the sufficiency part is almost the same as the proof
of that linear independency implies capability of exact cloning in
Ref. \cite{DG98}. To be complete, we outline here the main steps.

To show the existence of a desired separation under the assumption
of Eq.(\ref{large}), we need only to prove that there exists a
unitary transformation $U$ such that for any $i=1,\dots,n$,
\begin{equation}\label{unitary}
U|\psi_i\>_A|\Sigma\>_B|P\>_P=\sqrt{\gamma_i}|\psi'_i\>_{AB}|P_0\>_P+\sum_{k=1}^n
c_{ik}|\Phi_i\>_{AB}|P_k\>_P,
\end{equation}
where $|P_0\>,|P_1\>,\dots,|P_n\>$ are orthonormal states in the
probe system $P$, and $|\Phi_i\>_{AB}$ are normalized but not
necessarily orthogonal states. Here the subscript $B$ denotes an
ancillary system and $|\Sigma\>$ is a standard `blank' state (in
some cases, say unambiguous discrimination, the ancillary system
is unnecessary). After the unitary evolution described by
Eq.(\ref{unitary}), a projective measurement which consists of
$|P_0\>\<P_0|$ and $I-|P_0\>\<P_0|$ is performed on probe system
$P$. If the outcome corresponding to $I-|P_0\>\<P_0|$ occurs, the
separation fails; otherwise this separation succeeds and the
secretly chosen state $|\psi_i\>$ conclusively evolves into the
desired state $|\psi'_i\>$.

In the following, we show the existence of the unitary
transformation $U$ in Eq.(\ref{unitary}). Taking the inter-inner
products of the both sides of Eq.(\ref{unitary}) for different $i$
and $j$, we have the matrix equation
\begin{equation}\label{matrix}
X=\sqrt{\Gamma} X'\sqrt{\Gamma} + CC^\dag,
\end{equation}
where $n\times n$ matrix $C=[c_{ij}]$. From Lemma 1, the only
thing left is to show the existence of the matrix $C$. But from
Eq.(\ref{large}), the positive semidefinite matrix
$X-\sqrt{\Gamma}X'\sqrt{\Gamma}$ can be diagonalized by a unitary
matrix $V$ as
\begin{equation}
V(X-\sqrt{\Gamma}X'\sqrt{\Gamma})V^\dag=\diag(c_1,\dots,c_n)
\end{equation}
for some nonnegative numbers $c_1,\dots,c_n$. So we need only set
$C=V^\dag \diag(\sqrt{c_1},\dots,\sqrt{c_n})V^\dag$ and then the
sufficiency part of the theorem is proven.
 \hfill
$\blacksquare$

Theorem 1 tells us when a $given$ separation can be physically
realized in pure state case. The following theorem, however, gives
a necessary and sufficient condition under which $any$ quantum
separation is realizable on a given system in the general case of
mixed states. To begin with, we introduce some notations. For a
density matrix $\rho$, we denote by $\supp(\rho)$ the support
space of $\rho$. That is, the space spanned by all eigenvectors
with nonzero corresponding eigenvalues of $\rho$. Furthermore, by
$\supp(\rho_1,\dots,\rho_n)$ we denote the support space spanned
by eigenvectors of $\rho_1,\dots,\rho_n$ with nonzero
corresponding eigenvalues.

\begin{theorem}
Suppose a quantum system is prepared in a state secretly chosen
from $\rho_1,\dots,\rho_n$. Let $S=\{\rho_1,\dots,\rho_n\}$ and
$S_i=S\backslash \{\rho_i\}$. Then

1) any state separation on this system is possible (that is, for
any states $\rho'_1,\dots,\rho'_n$, there exists a separation
which leads $\rho_i$ conclusively to $\rho'_i$) if and only if
$\supp(S)\neq \supp(S_i)$ for any $i=1,\dots,n$.

2) Furthermore, if $\supp(S)= \supp(S_i)$ for some $i$ and there
exists a separation which leads $\rho_i$ conclusively to $\rho'_i$
for some quantum states $\rho'_1,\dots,\rho'_n$, then $\supp(S')=
\supp(S'_i)$, where  $S'=\{\rho'_1,\dots,\rho'_n\}$ and
$S'_i=S'\backslash \{\rho'_i\}$.
\end{theorem}

{\it Proof.} The necessity part of 1) is obvious, since we can
take special cases of quantum separation, say unambiguous
discrimination, to show that $\supp(S)\neq \supp(S_i)$ (for the
condition under which unambiguous discrimination between mixed
states is possible, we refer to Ref. \cite{FD04}).

To prove the sufficiency part of 1), suppose that $\supp(S)\neq
\supp(S_i)$ for any $i=1,\dots,n$. Then from Ref. \cite{FD04},
there exist $n$ positive real numbers $\gamma_1,\dots,\gamma_n$
such that we can unambiguously discriminate $\rho_i$ with
probability $\gamma_i$. Once the state $\rho_i$ is identified, we
can prepare $\rho'_i$ with certainty by a physical realizable
process (which may be dependent on $\rho'_i$). So by combining
these two steps together, we construct a protocol which leads
$\rho_i$ to $\rho'_i$ with positive probability $\gamma_i$.

Now we prove 2) by contradiction. Suppose $\supp(S')\neq
\supp(S'_i)$. Then there exists a pure state $|\phi\>$ which is in
$\supp(\rho_i)$ but not in $\supp(S'_i)$. So we can construct a
positive-operator valued measurement comprising $|\phi\>\<\phi|$
and $I-|\phi\>\<\phi|$ to unambiguously discriminate $\rho_i$ from
the other $n-1$ states with a positive probability. Notice that an
unambiguous discrimination is also a quantum separation. Combining
these two separation processes together we get a new one which can
discriminate unambiguously the state $\rho_i$ from other states
with a positive probability. That is a contradiction with the
assumption that $\supp(S)= \supp(S_i)$. \hfill $\blacksquare$

Notice that when
$\rho_1=|\psi_1\>\<\psi_1|,\dots,\rho_n=|\psi_n\>\<\psi_n|$ are
all pure states, the condition that $\supp(S)\neq \supp(S_i)$ for
any $i=1,\dots,n$ is equivalent to that
$|\psi_1\>,\dots,|\psi_n\>$ are linearly independent. So we have
the following corollary which has more physical intuition.

\begin{corollary}
Suppose a quantum system is prepared secretly in one of the states
$|\psi_1\>,\dots,|\psi_n\>$. Then

1) any state separation on this system is possible if and only if
$|\psi_1\>,\dots,|\psi_n\>$ are linearly independent.

2) Furthermore, if $|\psi_1\>,\dots,|\psi_n\>$ are linearly
dependent and there exists a separation which leads $|\psi_i\>$
conclusively to $|\psi'_i\>$ for some quantum states
$|\psi'_1\>,\dots,|\psi'_n\>$, then $|\psi'_1\>,\dots,|\psi'_n\>$
are also linearly dependent.
\end{corollary}

The two statements in Corollary 1 are complementary with each
other. Statement 2) tells us the constraints on realizable
information processing tasks when the system we are concerned with
is in a state coming secretly from a linearly dependent set. On
the other hand, statement 1) shows that linear dependency is
actually the only case in which physically realizable information
processing tasks will be constrained. That is, if the state of the
original system is prepared secretly in one of linearly
independent pure states, then any tasks, represented by our
generalized separation with arbitrary outcome states, are
probabilistically and conclusively realizable.

From Theorem 1, we get the following direct corollary:
\begin{corollary}
For any set $S=\{\rho_1,\dots,\rho_n\}$ of quantum states, the
following statements are equivalent:

1) The states secretly chosen from $S$ can be unambiguously
discriminated.

2) The states secretly chosen from $S$ can be conclusively cloned.

3) The set $S$ can evolve, through appropriate separation
processes, into any set $S'=\{\rho'_1,\dots,\rho'_n\}$ of quantum
states, where $\rho_i$ becomes $\rho'_i $ for any $i=1,\dots,n$.
\end{corollary}

Informally, from this corollary, exact cloning and unambiguous
discrimination put the strongest constraints on the possible
states the original system can be prepared in.

\section{Lower bound on average failure probability}

Theorem 1 gives a necessary and sufficient condition under which a
given separation can be realized for a given original system, when
the case of $pure$ state is considered. The general case where the
state of the system we are concerned with comes from a mixed state
set is, however, not investigated. Actually, it is unlikely that
there exists a corresponding condition for mixed states due to
lack of a result similar to Lemma 1. However, we can still derive
a lower bound on the average failure probability of any separation
once it is realizable.

\begin{theorem}
Suppose a quantum system is prepared in a state secretly chosen
from $\rho_1,\dots,\rho_n$ with respective $a$ $priori$
probabilities $\eta_1,\dots,\eta_n$, and there exists a separation
which leads $\rho_i$ to $\rho'_i$ for some quantum states
$\rho'_1,\dots,\rho'_n$. Then the average failure probability
$P_f$ of this separation satisfies
\begin{equation}\label{prob}
P_f\geq \sqrt{\frac{n}{n-1}\sum_{(i,j)\in \Delta}\eta_i
\eta_j\big(\frac{F(\rho_i,\rho_j)-F(\rho'_i,\rho'_j)}{1-F(\rho'_i,\rho'_j)}\big)^2},
\end{equation}
where the index set $\Delta=\{(i,j):i\neq j\ {\rm and}\
F(\rho'_i,\rho'_j)\leq F(\rho_i,\rho_j)\}$.
\end{theorem}

{\bf Proof.} From the assumption, there exist quantum operators
$A_{Sk}$ and $A_{Fk}$ satisfying the completeness relation
Eq.(\ref{identity}), such that Eqs.(\ref{1}) and (\ref{2}) hold.
It is easy to check that
\begin{equation}\label{Pf}
P_f=\sum_{i,k} \eta_i f_{ik}
\end{equation} and for any
$i=1,\dots,n$,
\begin{equation}\label{uni}
\sum_k (s_{ik}+f_{ik}) =1.
\end{equation}
By Cauchy-Schwarz inequality,
\begin{equation}\label{square}
\begin{array}{rl}
P_f^2 & \geq \displaystyle\frac{n}{n-1} \sum_{i\neq j} \eta_i
\eta_j\big(\sum_k
f_{ik}\big)\big(\sum_k f_{jk}\big)\\
\\
& \geq\displaystyle \frac{n}{n-1} \sum_{i\neq j} \eta_i
\eta_j\big(\sum_k \sqrt{f_{ik} f_{jk}}\big)^2.
\end{array}
\end{equation}

From Eq.(\ref{1}) and Polar decomposition theorem, we have
\begin{equation}
A_{Sk}\sqrt{\rho_i}=\sqrt{A_{Sk}\rho_i A_{Sk}^\dag} U_{ik}
=\sqrt{s_{ik}}\sqrt{\rho'_i}U_{ik}
\end{equation}
for some unitary matrix $U_{ik}$. And similarly, Eq.(\ref{2})
implies that
\begin{equation}
A_{Fk}\sqrt{\rho_i}=\sqrt{A_{Fk}\rho_i A_{Fk}^\dag} V_{ik}
=\sqrt{f_{ik}}\sqrt{\sigma_{ik}}V_{ik}
\end{equation}
for some unitary matrix $V_{ik}$.

Recall that for any density matrices $\rho$ and $\sigma$, the
fidelity $F(\rho,\sigma)=\max_U |\tr(\sqrt{\rho}\sqrt{\sigma}U)|$
, where the maximum is taken over all unitary matrix $U$. For any
$i\neq j$, let us take $U^j_{i}$ such that $F(\rho_i,\rho_j)=
|\tr(\sqrt{\rho_i}\sqrt{\rho_j}U^j_{i})|$. Then
\begin{equation}\label{Sk}
\tr(\sqrt{\rho_i}A_{Sk}^\dag
A_{Sk}\sqrt{\rho_j}U^j_{i})=\sqrt{s_{ik}s_{jk}}\tr(U_{ik}^\dag
\sqrt{\rho'_i}\sqrt{\rho'_j}U_{jk} U^j_{i})
\end{equation}
\begin{equation}\label{Fk}
\tr(\sqrt{\rho_i}A_{Fk}^\dag
A_{Fk}\sqrt{\rho_j}U^j_{i})=\sqrt{f_{ik}f_{jk}}\tr(V_{ik}^\dag
\sqrt{\sigma_{ik}} \sqrt{\sigma_{jk}}V_{jk} U^j_{i}).
\end{equation}
Summing up Eqs.(\ref{Sk}) and (\ref{Fk}) for all $k$ and noticing
Eq.(\ref{identity}), we have
\begin{equation}
\begin{array}{rl}
F(\rho_i,\rho_j)= & |\displaystyle\sum_k(\sqrt{s_{ik}s_{jk}}\tr(
\sqrt{\rho'_i}\sqrt{\rho'_j}W_{ijk})\\
\\
& +\displaystyle \sqrt{f_{ik}f_{jk}}\tr(\sqrt{\sigma_{ik}}
\sqrt{\sigma_{jk}}W'_{ijk}))|,
\end{array}
\end{equation}
where $W_{ijk}=U_{jk}U^j_{i}U_{ik}^\dag$ and
$W'_{ijk}=V_{jk}U^j_{i}V_{ik}^\dag$ are unitary matrices. We
further derive that
\begin{equation}\label{F}
\begin{array}{rll}
F(\rho_i,\rho_j) & \leq &
\displaystyle\sum_k\sqrt{s_{ik}s_{jk}}|\tr(
\sqrt{\rho'_i}\sqrt{\rho'_j}W_{ijk})|\\
\\
& &+ \displaystyle\sum_k
\sqrt{f_{ik}f_{jk}}|\tr(\sqrt{\sigma_{ik}}
\sqrt{\sigma_{jk}}W'_{ijk})| \\
\\
&\leq &\displaystyle\sum_k\sqrt{s_{ik}s_{jk}}F(\rho'_i,\rho'_j) \\
\\
& &+\displaystyle \sum_k
\sqrt{f_{ik}f_{jk}}F(\sigma_{ik},\sigma_{jk}) \\
\\
&\leq &\displaystyle\sum_k\sqrt{s_{ik}s_{jk}}F(\rho'_i,\rho'_j) +
\sum_k \sqrt{f_{ik}f_{jk}}.
\end{array}
\end{equation}
Notice that
\begin{equation}\label{each}
\begin{array}{rl}
\displaystyle\sum_k\sqrt{\displaystyle s_{ik}s_{jk}} & \leq
\displaystyle\sum_k
\frac{s_{ik}+s_{jk}}{2} \\
 \\
&=1- \displaystyle \sum_k \frac{f_{ik}+f_{jk}}{2}\\
 \\
&\leq 1- \displaystyle\sum_k\sqrt{f_{ik}f_{jk}}.
\end{array}
\end{equation}
Substituting Eq.(\ref{each}) into Eq.(\ref{F}), we have
\begin{equation}\label{k}
\sum_k\sqrt{f_{ik}f_{jk}}\geq
\frac{F(\rho_i,\rho_j)-F(\rho'_i,\rho'_j)}{1-F(\rho'_i,\rho'_j)}
\end{equation}
Taking Eq.(\ref{k}) for $(i,j)\in \Delta$ back into
Eq.(\ref{square}) and noticing that $\sum_k\sqrt{f_{ik}f_{jk}}\geq
0$ for $(i,j)\not\in \Delta$, we arrive at the desired bound,
\begin{equation}
P_f\geq \sqrt{\frac{n}{n-1}\sum_{(i,j)\in \Delta}\eta_i
\eta_j\big(\frac{F(\rho_i,\rho_j)-F(\rho'_i,\rho'_j)}{1-F(\rho'_i,\rho'_j)}\big)^2}.
\end{equation}
That completes the proof. \hfill $\blacksquare$

Following the argument behind Theorem 3 in Ref. \cite{FD04}, we
can derive a series of lower bounds on the average failure
probability. For the sake of completeness, we outline the
derivation as follows. Define

\begin{equation}M_t=\sum_i \eta_i^{2t} (\sum_k f_{ik})^{2t}\end{equation} and
\begin{equation}N_t=\sum_{i\not=j} \eta_i^t\eta_j^t (\sum_k f_{ik})^t(\sum_k
f_{jk})^t.\end{equation} Then $M_t=\sqrt{N_{2t}+M_{2t}}$ and by
Cauchy inequality, $M_{t}\geq N_t/(n-1)$. So for any $r\geq 0$,
\begin{equation}
\begin{array}{ll}
P_f^2&=N_1+M_1=N_1+\sqrt{N_2+M_2}=\cdots\\
\\
&=N_1+\sqrt{N_2+\sqrt{\dots+\sqrt{N_{2^r}+M_{2^r}}}}\\
\\
&\geq N_1+\sqrt{N_2+\sqrt{\cdots+\sqrt{\frac{n}{n-1}N_{2^r}}}}.
\end{array}
\end{equation}
If we further define
\begin{equation}
 C_t=
\sum_{(i,j)\in \Delta}\eta_i^t
\eta_j^t\big(\frac{F(\rho_i,\rho_j)-F(\rho'_i,\rho'_j)}{1-F(\rho'_i,\rho'_j)}\big)^{2t},
\end{equation} then from Eq.(\ref{k}) and the fact that
$\sum_k\sqrt{f_{ik}f_{jk}}\geq 0$ for $(i,j)\not\in \Delta$, we
have $N_t\geq C_t$. Consequently, the promised lower bounds on the
average failure probability $P_f$ can be derived as
\begin{equation}
P_f\geq P_f^{(r)}\doteq
\sqrt{C_1+\sqrt{\cdots+\sqrt{\frac{n}{n-1}C_{2^r}}}}.
\end{equation}
The bound presented in Eq.(\ref{prob}) is just the special case of
the above bounds when $r=0$. Note that $P_f^{(0)}\leq
P_f^{(1)}\leq \cdots$ by Cauchy-Schwarz inequality. When $r$
increases, the bound becomes better and better; and the limit when
$r$ tends to infinity is the best bound we can derive using this
method.

Now let us analyze the bound in Eq.(\ref{prob}) carefully. First,
note that when pure state separation is considered, Qiu obtained
in Ref. \cite{QI02} a lower bound on the average failure
probability which reads
\begin{equation}\label{qiubound}
1-\frac{1}{n-1}\sum_{i<j}\frac{\eta_i+\eta_j-2\sqrt{\eta_i\eta_j}|\<\psi_i|\psi_j\>|}{1-|\<\psi'_i|\psi'_j\>|}.
\end{equation}
It is easy by using Cauchy-Schwarz inequality to check that our
bound presented in Eq.(\ref{prob}) is better in general than the
one in Eq.(\ref{qiubound}). On the other hand, in the case of
$M\rightarrow N$ ($M\leq N$) exact cloning, where the original
state and the final state are, respectively, $\rho_i^{\otimes M}$
and $\rho_i^{\otimes N}$ for $i=1,\dots,n$, and so
$F(\rho'_i,\rho'_j)\leq F(\rho_i,\rho_j)$ holds for any $i\neq j$.
So we have actually derived a lower bound on the average failure
probability of exact $M\rightarrow N$ cloning as
\begin{equation}
P_f^{EC}\geq \sqrt{\frac{n}{n-1}\sum_{i\neq j}\eta_i
\eta_j\big(\frac{F(\rho_i,\rho_j)^M-F(\rho_i,\rho_j)^N}{1-F(\rho_i,\rho_j)^N}\big)^2}.
\end{equation}
When $\rho_i=|\psi_i\>\<\psi_i|$ are pure states and
$\eta_1=\dots=\eta_n=1/n$, this bound can be shown better than
\begin{equation}
1-\frac{2}{n(n-1)}\sum_{i<j}\frac{1-|\<\psi_i|\psi_j\>|^M}{1-|\<\psi_i|\psi_j\>|^N},
\end{equation}
which was derived in Ref. \cite{CB98}. Finally, in the case of
unambiguous discrimination, where the final states $\rho'_i$ are
orthogonal to each other, the bound in Eq.(\ref{prob}) turns out
to be
\begin{equation}
P^{UD}_f\geq \sqrt{\frac{n}{n-1}\sum_{i\neq j}\eta_i \eta_j
F(\rho_i,\rho_j)^2},
\end{equation}
coinciding with that obtained in Ref. \cite{FD04}. It is also
worth noting that the bound can further degenerate to the
Jaeger-Shimony bound $1-2\sqrt{\eta_1 \eta_2}|\<\psi_1|\psi_2\>|$
for two pure states \cite{JS95} and the IDP bound
$1-|\<\psi_1|\psi_2\>|$ for two pure states with equal $a$
$priori$ probabilities \cite{IV87,DI88,PE88}.

\section{Conclusion}
To conclude, by deriving a necessary and sufficient condition for
any quantum separation to be physically realizable, we show that
in probabilistic manner, linearity is in fact the only one that
restricts the physically realizable tasks. That is, when a system
is prepared in a state secretly chosen from a linearly independent
pure state set, then any generalized state separation is
physically realizable with a positive probability. A lower bound
on the average failure probability of any quantum state separation
is also derived and special cases of this bound are analyzed.

The authors thank the colleagues in the Quantum Computation and
Quantum Information Research Group for useful discussion. This
work was partly supported by the Natural Science Foundation of
China (Grant Nos. 60273003, 60433050, and 60305005). R. Duan
acknowledges the financial support of Tsinghua University (Grant
No. 052420003).

\end{document}